\documentstyle[preprint,aps,eqsecnum,prd]{revtex}

\begin{document}
\draft

\title{\vskip-3cm{\baselineskip14pt
\centerline{\normalsize\hskip12.5cm hep-ph/9412311}
\centerline{\normalsize\hskip12.5cm MPI/PhT/94--91}
\centerline{\normalsize\hskip12.5cm MAD/TH/94--1}
\centerline{\normalsize\hskip12.5cm TUM--HEP--200/94}
\centerline{\normalsize\hskip12.5cm December 1994}}
\vskip1.5cm
Two-loop ${\rm O}\left(G_F^2M_H^4\right)$ corrections to the
fermionic decay rates of the Higgs boson}
\author{Loyal Durand,\thanks{Electronic address:
ldurand@wishep.physics.wisc.edu}\,$^1$
Bernd A. Kniehl,\thanks{Electronic address: kniehl@mpiw16.mppmu.mpg.de}\,$^2$
and
Kurt Riesselmann\thanks{Electronic address: kurtr@physik.tu-muenchen.de}
\,$^3$}

\address{$^1$\,Department of Physics, University of Wisconsin,\\
1150 University Avenue, Madison, WI 53706, USA\\
$^2$\,Max-Planck-Institut f\"ur Physik, Werner-Heisenberg-Institut,\\
F\"ohringer Ring 6, 80805 M\"unchen, Germany\\
$^3$\,Physik-Department T30, Technische Universit\"at M\"unchen,\\
James-Franck-Stra\ss e, 85747 Garching b.\ M\"unchen, Germany}

\date{\today}
\maketitle

\begin{abstract}
We calculate the dominant ${\rm O}\left(G_F^2M_H^4\right)$ two-loop
electroweak corrections to the fermi\-onic decay widths of a
heavy Higgs boson in the Standard Model.
Use of the Goldstone-boson equivalence theorem reduces the problem
to one involving only the physical Higgs boson $H$ and the
Goldstone bosons $w^\pm$ and $z$ of the unbroken theory.
The two-loop corrections are opposite in sign to the one-loop
electroweak corrections, exceed
the one-loop corrections in magnitude for
$M_H>1114\ {\rm GeV}$, and increase in relative magnitude
as $M_H^2$ for larger values of $M_H$.
We conclude that the perturbation expansion in
powers of $G_FM_H^2$
breaks down for $M_H\approx 1100\ {\rm GeV}$.
We discuss briefly the QCD and
the complete one-loop electroweak corrections to $H\rightarrow b\bar{b},
\,t\bar{t}$, and comment on the validity of the equivalence theorem.
Finally we note how a
very heavy Higgs boson could be described in a phenomenological
manner.
\end{abstract}

\narrowtext

\section{INTRODUCTION}

One of the great puzzles of contemporary elementary particle research
is whether nature makes use of the Higgs mechanism to generate
the observed particle masses.
In the minimal standard model (SM)  of electroweak interactions,
the symmetry breaking
is implemented using this mechanism with one weak-isospin doublet of complex
scalar fields with weak hypercharge $Y=1$.
With the spontaneous breaking of the SU(2)$_L\otimes$U(1)$_Y$ gauge
symmetry, three of the four scalar degrees of freedom are absorbed to create
the longitudinal polarization states of the intermediate bosons, $W^\pm$
and $Z$.
At the same time, the quarks and charged leptons acquire masses through
their Yukawa interaction with the scalar doublet.
There remains at the end
one neutral scalar boson with positive parity and charge
conjugation, the physical Higgs boson $H$.

Most of the properties of the scalar or Higgs sector of the SM are fixed
experimentally, e.g., the vacuum expectation value,
$v=2^{-1/4}G_F^{-1/2}\approx 246 \ {\rm GeV}$,
the coupling of the Higgs boson to the gauge bosons,
$g_{VVH}=2^{5/4}G_F^{1/2}M_V^2$ where $V=W^{\pm},Z$,
and the coupling of the Higgs to the fermions,
$g_{f\bar fH}=2^{1/4}G_F^{1/2}m_f$.
However, the mass $M_H$ of the Higgs boson and its quartic self-coupling,
$\lambda=G_FM_H^2/\sqrt{2}$, are unspecified.
It is therefore of considerable interest
to analyze processes which can give theoretical limits on $M_H$, or test
the effects of the quartic coupling phenomenologically.

The range of possible Higgs masses is constrained
from below both experimentally and theoretically.
The non-detection of the $Z$-boson decay $Z\rightarrow f\bar f H$ at LEP~1
and SLC has ruled out a Higgs mass of less than 63.9~GeV at the 95\%
confidence level \cite{exp}. Depending on the mass of the top quark, the
requirement that the vacuum be the true ground state could provide an
even more stringent theoretical lower bound \cite{lin}.
Other theoretical arguments bound the Higgs mass from above.
Nonperturbative lattice computations \cite{has,neu} give an
upper limit for $M_H$ of about 710~GeV \cite{neu}.
Unitarity arguments in intermediate-boson scattering
at high energies \cite{dic,lee} and considerations concerning the
range of validity of
perturbation theory \cite{vel,mve} establish an upper bound $M_H<
(8\pi\sqrt2/3G_F)^{1/2}\approx 1\ {\rm TeV}$ in a weakly interacting theory.
The unitarity bound on $M_H$
is lowered significantly when the approach of \cite{dic,lee} is
extended to higher orders; see \cite{arg,unit} and
references therein.
However, the improved bound depends on the energy scale up to
which the SM is assumed to remain valid.

A violation of the unitarity bound on $M_H$ is presumably a
signal for the onset of strong
interactions in the Higgs sector of the SM, a possibility which is
of considerable interest in its own right \cite{strong}.
It would therefore be desirable to
sharpen the bound by removing the uncertainty associated with the mass
scale at which it is applied, or to find a separate
scale-independent bound. In fact,
the work presented here on two-loop electroweak corrections to the fermionic
decay modes of the Higgs boson, $H\rightarrow f\bar f$,
gives a scale-independent limit on $M_H$ in a weakly
interacting theory.
We find that the two-loop corrections to the
fermionic decay rates exceed the one-loop corrections in magnitude for
$M_H\approx 1114 \ {\rm GeV}$, and increase in relative magnitude
proportionally to $M_H^2$ for larger Higgs-boson masses. We conclude, as
reported previously \cite{prl,ghi,adrian},
that the perturbative expansion fails to converge satisfactorily, and that the
theory becomes effectively strongly interacting in the Higgs sector for
$M_H\gtrsim 1100 \ {\rm GeV}$.

The result noted above is a consequence of
our calculation of the dominant two-loop electroweak corrections to the
fermionic decay rates of the Higgs boson, the subject of this paper.
The fermionic decay modes are of considerable
phenomenological interest. For example, the Higgs boson decays predominantly
to $b\bar b$ pairs if $M_H\lesssim135 \ {\rm GeV}$.
The search for a low- or intermediate-mass Higgs boson at future high-energy
$e^+e^-$ linear colliders \cite{imh}, the Fermilab Tevatron \cite{msw},
or a possible 4-TeV upgrade thereof \cite{gun} will rely largely on this mode
by tagging the $B$ mesons.
Moreover, in this mass range, the branching fractions of all other decay
channels depend sensitively on the $H\rightarrow b\bar b$ decay width.
It has been argued that the low-mass Higgs boson might also be detectable
at future hadron supercolliders through the
$H\rightarrow\tau^+\tau^-$ signal \cite{rke}, while the decay
$H\rightarrow t\bar t$ will have an appreciable branching
fraction for $M_H>2m_t$.
Future high-energy $e^+e^-$ colliders will also be able to measure
the $Ht\bar t$ Yukawa coupling \cite{hag}.

The measurement of the Higgs-boson mass and couplings in future experiments
will require an understanding
of the radiative corrections to the fermionic decay rates of the Higgs
boson.  Much work has been done in this area, and a recent review is
given in \cite{kni}. There are important differences between the
radiative corrections involved in Higgs physics, and those familiar in the
gauge sector of the SM, e.g., in $Z$-boson decays.
It is well known that the quantum effects induced by
virtual Higgs bosons are {\it screened} in $Z$-boson physics
\cite{vel,lem}: they depend only logarithmically on $M_H$ at one
loop, and are quadratic in $M_H$, but with minute coefficients,
at two loops \cite{hal}.
In contrast, the one-loop electroweak corrections to the partial decay widths
\cite{vel,mar,hvv,hff} and production cross sections \cite{pro}
of the Higgs boson
are already dominated for $M_H\gg M_W$ by terms quadratic in $M_H$.
These terms give rise to moderate enhancements
of the rates for Higgs masses of up to $1 \ {\rm TeV}$.
However, it is premature to conclude that the
two-loop electroweak corrections will
also be perturbatively small in the high-$M_H$ range, since these
corrections have terms {\it quartic} in $M_H$.
It is of both theoretical and phenomenological interest to check
the importance of these potentially large corrections by explicit calculation.

In this paper, we calculate the complete ${\rm O}\left(G_F^2M_H^4\right)$
corrections to the fermionic decay
rates of a Higgs boson with $M_H\gg M_W$. These corrections,
which are the leading two-loop electroweak corrections for $M_H\gg M_W$,
are independent of the fermion flavor, and, as noted above, are larger
than the one-loop corrections of ${\rm O}\left(G_FM_H^2\right)$ for
$M_H>1114\ {\rm GeV}$.
We compare our results with other known one-loop corrections.
To obtain the full electroweak  two-loop
corrections for specific fermion channels in the limit $M_H\gg M_W$,
one would have to calculate further,
flavor-dependent corrections of mixed orders in the Higgs and Yukawa couplings,
namely ${\rm O}\left( G_F^2M_H^2m_f^2\right)$
and ${\rm O}\left(G_F^2m_f^4\right)$.
These corrections, however, are not universal. For example,
different fermionic channels,
such as $H\rightarrow \tau^+\tau^-,\ b\bar b,\
t\bar t$, all have different dependence on $m_t$.

\section{CALCULATION OF THE ${\rm O}\left(G_F^2M_H^4\right)$ CORRECTIONS}

In this section, we sketch the calculation of the dominant, flavor-independent
electroweak corrections to the decay rate $H\rightarrow f\bar f$.
The starting point of our analysis is the bare Lagrangian for the
Higgs-fermion interaction,

\begin{equation}
{\cal L}_{f,0}^{\rm Yuk}=-{m_{f,0}\over v_0}\,\bar\psi_{f,0}
H_0\psi_{f,0}\,,
\label{eqyuk}
\end{equation}

\noindent where the subscript ``0'' denotes bare quantities.
Our aim is to obtain the leading, flavor-independent corrections to
the $Hf\bar{f}$ vertex in powers of $G_FM_H^2$. The mass and wave-function
renormalization constants for the fermions as well as the loop corrections
to the vertex depend on the Yukawa couplings for the fermions,
and are omitted consistently because of their subleading nature.
(We will return to these corrections at the
one-loop level in Sec.\ IIIB.)
We may therefore replace the fermionic quantities
$m_{f,0}$ and $\psi_{f,0}$ in Eq.\ (\ref{eqyuk})
by $m_f$ and $\psi_f$. The contributions to the
renormalization constants for the bare Higgs
field, $H_0$, and the bare vacuum expectation value, $v_0$,
in powers of $G_FM_H^2$
are determined entirely by the symmetry-breaking sector of the
SM. The contributions of fermion loops to these two quantities
depend on the Yukawa couplings
and are again omitted as they do not contribute to the ${\rm O}(G_F^2M_H^4)$
corrections (see Sec.\ IIIB). We may therefore
calculate the desired corrections to the decay $H\rightarrow f\bar{f}$ vertex
with all Yukawa couplings set to zero.

The calculation can be simplified greatly in the limit of interest,
$M_H\gg M_W$, through the use
of the Goldstone-boson equivalence theorem \cite{cor}. This theorem
states that the leading electroweak contribution to a graph in powers
of $G_FM_H^2$ can be calculated by replacing the gauge bosons $W^\pm$,
$Z$ by the would-be Goldstone bosons $w^\pm$, $z$ of the symmetry-breaking
sector of the theory.
Because $M_W/M_H\propto gv/M_H$, we can simplify our calculation
consistently in the limit of a heavy Higgs boson
by neglecting the gauge couplings $g,g'$ and
taking the Goldstone bosons to be massless.
Adopting the conventions of \cite{mah}, we can
write the relevant Lagrangian for the symmetry-breaking sector of the SM
in terms of bare quantities as follows:

\begin{eqnarray}
{\cal L}_0^{\rm SBS}&=&\frac{1}{2}\partial_\mu{\bf w}_0\cdot\partial^\mu
{\bf w}_0+\frac{1}{2}\partial_\mu H_0\,\partial^\mu H_0
-\frac{1}{2}M_{w,0}^2 {\bf w}_0^2-\frac{1}{2}M_{H,0}^2 H_0^2 \label{eqlag} \\
& &-{\lambda_0\over4}\left({\bf w}_0^2+H_0^2\right)^2
-\lambda_0 v_0\left({\bf w}_0^2+H_0^2\right)H_0\,,\nonumber
\end{eqnarray}

\noindent where the real scalar triplet,
${\bf w}=(w_1,w_2,w_3)$, is related to the
Goldstone bosons, $w^\pm$ and $z$, by
$w^\pm=({w}_1\mp i{w}_2)/\sqrt{2}$ and $z={w}_3$, respectively.
The tadpole counterterm, which cancels all tadpole contributions of
${\cal L}_0^{\rm SBS}$ order by order,
has been omitted in writing Eq.\ (\ref{eqlag}); therefore all
graphs which include tadpole contributions need to be dropped in calculations
\cite{mah}. Note that a full equivalence-theorem calculation would also require
the complete Yukawa Lagrangian for the interactions of fermions with the
massless Goldstone bosons and the Higgs boson \cite{ldkr}. Because we are not
interested in corrections due to the Yukawa couplings we do not give the
complete Yukawa Lagrangian here, except for the piece given in Eq.\ (2.1).

The on-mass-shell renormalization is carried out in such a way
that the physical
mass of the Higgs boson, defined in terms of the position of
the pole in the Higgs propagator, is
$M_H$. The three Goldstone bosons remain massless to all orders in the
perturbation expansion in $\lambda_0$, and satisfy a residual SO(3)
symmetry \cite{breaking}. Requiring that
the residues of the physical on-shell propagators be unity fixes the
wave-function renormalization constants defined by the relations
$w_0^\pm=Z_w^{1/2}w^\pm$, $z_0=Z_z^{1/2}z$, and $H_0=Z_H^{1/2}H$,
where $w^\pm$, $z$, and $H$ are the physical fields.
The result is \cite{mah}

\begin{eqnarray}
{1\over Z_w}&=&1-\left.{d\over dp^2}
\Pi_w^0(p^2)\right|_{p^2=0}\,,\label{eqrenorm}\\
{1\over Z_z}&=&1-\left.{d\over dp^2}
\Pi_z^0(p^2)\right|_{p^2=0}\,,\nonumber \\
{1\over Z_H}&=&1-\left.{d\over dp^2}{\rm Re}
\Pi_H^0(p^2)\right|_{p^2=M_H^2}\,,\nonumber
\end{eqnarray}

\noindent where $\Pi_w^0(p^2)$, $\Pi_z^0(p^2)$, and $\Pi_H^0(p^2)$ are
the self-energy functions for
the bare fields calculated from the Lagrangian in Eq.\ (\ref{eqlag}).
Because of the SO(3) symmetry, $\Pi_w^0=\Pi_z^0$ and $Z_w=Z_z$.
Explicit expressions for the $\Pi^0$'s correct to two loops
may be found in Eqs.~(11) and (12) of \cite{mah}.
Furthermore, we have \cite{mah}

\begin{eqnarray}
M_{H,0}^2&=&M_H^2-{\rm Re}\Pi_H^0\left(M_H^2\right)\,,\\
M_{w,0}^2&=&-{\rm Re}\Pi_w^0(0)=-\Pi_w^0(0)\,,\nonumber\\
v_0&=&Z_w^{1/2}v\,,\nonumber\\
\lambda_0&=&{\lambda\over Z_w}\left(1-
{{\rm Re}\Pi_H^0\left(M_H^2\right)-\Pi_w^0(0)\over M_H^2}\right)\,.\nonumber
\end{eqnarray}

Using these results, we can write the Lagrangians above
entirely in terms of
renormalized, physical quantities.
The physical vacuum expectation value is fixed in terms of the
Fermi constant $G_F$ by the familiar relation
$v=2^{-1/4}G_F^{-1/2}$, while the physical quartic coupling is given by
$\lambda=G_FM_H^2/\sqrt{2}$.
The renormalized symmetry-breaking Lagrangian is given in
Eq.~(7) of \cite{mah}, while the renormalized form of the Higgs-fermion
Lagrangian is given for our purposes by

\begin{equation}
{\cal L}^{\rm Yuk}_f=-{{Z_H^{1/2}}\over Z_w^{1/2}}\,{m_f\over v}\,
\bar\psi_f H\psi_f\,.
\end{equation}

\noindent The radiatively corrected fermionic decay rate
of the Higgs boson is consequently given by

\begin{equation}
\Gamma\left(H\rightarrow f\bar f\,\right) = {{Z_H}\over {Z_w}}\,
\Gamma _B\left(H\rightarrow f\bar f\,\right)\,,
\label{eqcor}
\end{equation}

\noindent where \cite{born}

\begin{equation}
\Gamma _B\left(H\rightarrow f\bar f\,\right)=
{N_cm_f^2M_H\over8\pi v^2}\left(1-{{4{m_f^2}}\over {M_H^2}}\right)^{3/2}
\label{eqborn}
\end{equation}

\noindent is the Born result.
Here $N_c=1$ (3) for lepton (quark) flavors.

The wave-function renormalization constants
$Z_H$ and $Z_w$ were calculated to two loops,
${\rm O}\left(G_F^2M_H^4\right)$, in \cite{mah} using dimensional
regularization.
The two-loop diagrams that contribute to $Z_H$ and $Z_w$ at that order through
the derivatives of the self-energy functions in Eq.\ (\ref{eqrenorm})
are shown in Fig.\ \ref{diagrams};
no single diagram gives an exceptionally large
contribution. The results of the calculation
can be written in the form

\begin{equation}
{1\over Z_\sigma}=1+\hat{\lambda}\xi^\epsilon
\left(\mbox{\rule{0mm}{4.5mm}}a_\sigma +{\rm O}(\epsilon)\right)
+\hat{\lambda}^2\xi^{2\epsilon}\left({3\over \epsilon}+ b_\sigma
+{\rm O}(\epsilon)\right)
+{\rm O}\!\left(\hat{\lambda} ^3\right)
\qquad (\sigma = w, H)\,,\label{eqzw}\\
\end{equation}

\noindent where $\hat{\lambda}=(\lambda/16\pi^2)$, $\xi=4\pi\mu^2/M_H^2$,
$\epsilon=(4-D)/2$, $D$ is the dimensionality of space-time, and
$\mu$ is the arbitrary scale parameter introduced in the interaction to
keep $\lambda$ dimensionless for $\epsilon\not=0$. The coefficients
in the expansions above are:

\begin{eqnarray}
a_w&=&1\,,\label{eqcoeff}\\
b_w&=& \frac{3}{2} + 2\zeta(2)  - 6\gamma - 3\pi\sqrt{3}
       + 12{\rm Cl_2\left({\pi\over3}\right)}\sqrt{3} \nonumber\\
&\approx& 6.098\,, \nonumber\\
a_H&=& -12 + 2\pi\sqrt{3}\nonumber\\
&\approx&-1.12\,, \nonumber\\
b_H&=& \frac{291}{2}  - 96\zeta(2) + 90\zeta(3) - 6\gamma
       - 48\pi{\rm Cl_2\left({\pi\over3}\right)}
 + 116\pi\sqrt{3} - 216{\rm Cl_2\left({\pi\over3}\right)}\sqrt{3} - 162K_5
 \nonumber\\
&\approx& 41.12 \,. \nonumber
\end{eqnarray}

\noindent The constant $K_5=0.92363\ldots$
was evaluated numerically from Eq.~(A86) of \cite{mah}.
The Riemann $\zeta$ function takes the values
$\zeta(2)=\pi^2/6$ and $\zeta(3)=1.20205\ldots$,
${\rm Cl}_2$ is Clausen's function, ${\rm Cl}_2({\pi\over3})=1.01494\ldots$,
and $\gamma=0.57721\ldots$ is Euler's constant.
The one-loop coefficients $a_H$ and $a_w$ are similar in magnitude, but the
two-loop coefficients $b_H$ and $b_w$ differ in magnitude by roughly a
factor of 7, despite the fact that almost the same number of diagrams,
with similar structures and magnitudes, contribute.
It is also interesting that the coefficients in $Z_H^{-1}$ alternate
in sign; those in $Z_w^{-1}$ do not. We note that the results above,
revised relative to our previous analysis \cite{prl},
are now in complete agreement with the those of Ghinculov
\cite{ghi}, which have also been revised \cite{adrian}.

Because the decay width $\Gamma\left(H\rightarrow f\bar f\right)$
is a physical quantity, and
all radiative corrections that depend only on $G_FM_H^2$ are
contained in the factor $Z_H/Z_w$ in Eq.\ (2.6), this factor must be finite
for $\epsilon\rightarrow 0$. The $Z$'s are finite at one loop,
but not at two loops. Hence, the parts of the two-loop contributions to
$Z_H$ and $Z_w$ that are proportional to $1/\epsilon$
must cancel in the ratio.
The cancellation is clear if the $Z$'s are written in factored form,

\begin{equation}
 {1\over Z_\sigma}=
 \left( 1 + a_\sigma\hat{\lambda}\xi^\epsilon
+ b_\sigma\hat{\lambda}^2\xi^{2\epsilon}  \right)
\left( 1 + {3\over\epsilon}\hat{\lambda}^2\xi
^{2\epsilon}\right)
+{\rm O}\left(\hat{\lambda}^3\right) \qquad (\sigma = w, H)\,,
\label{eqzwf}
\end{equation}

\noindent and is exact to all orders in $\lambda$.
The complete cancellation of the divergent
terms allows us to take the limit $\epsilon \rightarrow 0$. In this limit,
$\xi^{\epsilon} \rightarrow 1$ with no pieces left over,
and the final ratio is independent of the scale $\mu$ introduced in the process
of dimensional regularization.

The ${\rm O}\left(G_F^2M_H^4\right)$  electroweak corrections to the
fermi\-onic decay rates emerge naturally in this formalism as the finite ratio

\begin{equation}
{{Z_H}\over {Z_w}} = { {1 + a_w\hat{\lambda} + b_w\hat{\lambda}^2}\over
                        {1 + a_H \hat{\lambda} + b_H \hat{\lambda}^2} }\,.
\label{eqres}
\end{equation}

\noindent This expression for $Z_H/Z_w$
automatically resums one-particle-reducible Higgs-boson self-energy
diagrams in a way that conforms
with the standard procedure
in $Z$-boson physics; see, e.g., \cite{bee}. However, it
is clear that the resummation contains
only limited information on higher-order terms.
Since we actually have no control of terms beyond
${\rm O}\left(\hat{\lambda}^2\right)$, and are not aware of a physical
principle which would select this as an optimum resummation scheme, we
expand Eq.\ (\ref{eqres}) and discard terms beyond
${\rm O}\left(\hat{\lambda}^2\right) = {\rm O}\left(G_F^2M_H^4\right)$.
This gives the alternative representation

\begin{eqnarray}
\label{percent}
{{Z_H}\over {Z_w}} &=& 1+(a_w-a_H)\hat{\lambda}
+\left(b_w - b_H - a_wa_H + a_H^2\right)\hat{\lambda}^2 \label{eqexp}\\
 &\approx& 1 + 2.12\hat{\lambda} - 32.66\hat{\lambda}^2 \nonumber\\
 &\approx& 1 + 0.013{\lambda} - 0.0013{\lambda}^2 \nonumber\\
&\approx&1+11.1\%\left({M_H\over1\,{\rm TeV}}\right)^2
- 8.9\%\left({M_H\over1\,{\rm TeV}}\right)^4\,.\nonumber
\end{eqnarray}

\noindent The result agrees at ${\rm O}\left(\hat{\lambda}\right)$
with the known one-loop result \cite{vel,mar},

\begin{equation}
{Z_H\over Z_w}=1+{G_FM_H^2\over8\pi^2\sqrt2}\left({13\over2}-\pi\sqrt3\,\right)
\,.
\end{equation}

\section{RESULTS}
\subsection{Limits on perturbation theory}
We are now in a position to explore the phenomenological implications
of our results.
In Fig.\ \ref{hff_ratio}, we show the leading electroweak
corrections to $\Gamma\left(H\rightarrow f\bar f\,\right)$ in the one- and
two-loop approximations with and without resummation of
one-particle-reducible higher-order terms plotted as functions of $M_H$.
We will concentrate first on the expanded results given in Eq.\ (\ref{eqexp}).
While the ${\rm O}\left(G_F^2M_H^4\right)$ term
(upper solid line in Fig.\ \ref{hff_ratio}) gives
a modest increase of the rates, e.g., by 11\% at $M_H=1 \ {\rm TeV}$,
the situation changes  when the two-loop term is included.
The importance of this term, which grows as $M_H^4$, increases
with $M_H$ in such a way that it cancels the one-loop term completely for
$M_H=1114 \ {\rm GeV}$, and is twice the size of the one-loop term, with the
opposite sign, for $M_H=1575 \ {\rm GeV}$. The total correction, shown by the
lower solid line in Fig.\ \ref{hff_ratio}, is then negative and has the same
magnitude as the one-loop correction alone. The perturbation series for the
corrections to
$\Gamma\left(H\rightarrow f\bar f\,\right)$ clearly ceases to converge
usefully, if at all, for $M_H\approx 1100\ {\rm GeV}$, or
equivalently, for $\lambda\approx 10$.
A Higgs boson with a mass larger than about 1100 GeV
effectively becomes a strongly
interacting particle.
Conversely, $M_H$ must not exceed approximately 1100~GeV if the standard
electroweak perturbation theory is to be predictive for the decays
$H\rightarrow f\bar{f}$.
Note that one cannot use the usual unitarization
schemes invoked in studies of $W_L^\pm,Z_L,H$ scattering \cite{lee,uni}
to restore the predictiveness for the heavy-Higgs width, as no
unitarity violation is involved.

One might expect to improve the perturbative result
in the upper range of $M_H$ somewhat
by resumming the one-particle-reducible contributions to the Higgs-boson
wave-function renormalization by using Eq.\ (\ref{eqres}) rather than
Eq.\ (\ref{eqexp}).
This leads to an increase of the one-loop correction
(upper dotted line in Fig.\ \ref{hff_ratio}),
while the negative effect of the two-loop correction
is lessened (lower dotted line) for large values of $M_H$.
However, in the mass range below $M_H=1400 \ {\rm GeV}$,
this effect is too small
to change our conclusions concerning the breakdown of perturbation theory.
Moreover, the resummed expression for the one-loop terms in the perturbation
expansion, when reexpanded to ${\rm O}\left(G_F^2M_H^4\right)$,
does not yield a proper estimate
for the size of the two-loop terms. There is consequently no reason to
favor this approach to the present problem.

It might be argued that the apparent breakdown in the perturbation
expansion as judged by a comparison of the one- and two-loop terms
is an artifact of a small one-loop contribution rather than
a consequence of large two-loop terms. However,
we see no evidence in the calculation that there are unusual cancellations
in the one-loop corrections. In fact, the one-loop contributions  $a_H$
and $a_w$  add
in magnitude in the
ratio $Z_H/Z_w$, whereas the two-loop contribution $b_w$ is in magnitude
subtracted from $b_H$ [see Eqs.\ (\ref{eqcoeff}) and (\ref{eqexp})].

In a previous publication, we looked at other processes to which $Z_w$
and $Z_H$ contribute, e.g., the scattering of Higgs bosons and
longitudinally polarized $W^\pm$ and $Z$ bosons \cite{unit}.
In the latter case, even
larger coefficients  appear in the perturbation expansion when it is
expressed, as above, in a series in the (running)
parameter $\hat{\lambda}_{\rm s}=(\lambda_{\rm s}/16\pi^2)$
\cite{a_1}. The contributions due to the finite parts of $Z_H$ and $Z_w$ are
rather insignificant in comparison with the finite parts of the
unrenormalized two-loop scattering graphs.

While
the factors of $(1/16\pi^2)$ occur naturally at each order in
perturbation theory, their incorporation into the {\it natural} parameter
$\hat{\lambda}$ is misleading: the coefficients of the \mbox{zero-,}
\mbox{one-,} and two-loop
terms in the diagonal partial-wave scattering amplitudes, and the one-
and two-loop terms in the Higgs-boson decay rate calculated above, all
have similar magnitudes when the series are
rewritten as expansions in the physical parameters
$\lambda_{\rm s}$ and $\lambda=G_FM_H^2/\sqrt{2}$
as seen in Eq.\ (\ref{percent}) and the comment \cite{a_1}.
It appears, therefore, that $\lambda$, and not
$\hat{\lambda}=\lambda/(16\pi^2)$, is
the natural expansion
parameter.

The high-energy scattering processes give strong evidence for a breakdown of
the perturbation series for a running coupling
$\lambda_{\rm s}(\sqrt{s})\approx 2.3$
at either the one-loop \cite{arg} or two-loop \cite{unit} level. This
translates to $M_H\approx 380\ {\rm GeV}$
if the SM is {\em assumed} to remain valid for energies $\sqrt{s}$
up to $\sim 5$ TeV \cite{unit}. The $M_H$ upper bound obtained
in the present analysis is considerably less stringent than the one found in
\cite{unit}.
However, we emphasize that the result obtained here does not
depend on the extra assumption about an energy scale;
the breakdown of perturbation theory is fixed solely
by the physical value of $M_H$.

\subsection{Comparison with complete one-loop corrections}

The leading corrections discussed so far are independent of the flavor of
the final-state fermion. However,
from an experimental point of view, they are relevant only for the
$t\bar{t}$ and, perhaps, the $b\bar{b}$ and $\tau^+\tau^-$
decays of the Higgs boson.
It is therefore interesting to compare
the corrections calculated here with the full one-loop
electroweak corrections \cite{hff}, and the QCD corrections that are
available for these decay channels \cite{bra,gor}.
In particular, the subleading two-loop electroweak corrections, those of
${\rm O}\left(G_F^2M_H^2m_t^2\right)$
and ${\rm O}\left(G_F^2m_t^4\right)$, are still unknown, but one may
estimate their likely importance by comparing the
top-quark Yukawa-coupling correction
to the Higgs-coupling correction at one loop.
The QCD corrections to the $H\rightarrow q\bar q$ modes,
where $q$ denotes a quark flavor, are known to ${\rm O}(\alpha_{\rm s})$ for
arbitrary values of $m_q$ \cite{bra} and
to ${\rm O}\left(\alpha_{\rm s}^2\right)$
in the limit $m_q\ll M_H$ \cite{gor}.
Their main effect is to replace the pole mass of
the quark $q$ by its $\overline{\rm MS}$ mass evaluated at the scale $M_H$.
For completeness, we mention that the
${\rm O}\left(\alpha_{\rm s} G_Fm_t^2\right)$
corrections to the fermionic decay rates are also now
available \cite{ks}.

The results in the preceding section were derived using the
Goldstone-boson equivalence theorem and neglecting further electroweak
corrections of orders $g^2$, $\lambda g^2$, etc., as well as contributions
that involve the Yukawa and QCD couplings of the fermions. Since $\lambda$,
$g^2$, $\alpha_{\rm s}$, and the Yukawa couplings are independent
parameters of the theory, our conclusion that the perturbation series in
$\lambda=G_FM_H^2/\sqrt{2}$ fails to converge satisfactorily is
independent of further corrections involving powers of $g$ and the other
independent couplings, though those further corrections may be numerically
important in applications of the results.

The use of the equivalence theorem
provides a correct framework for calculating the leading electroweak
corrections---those enhanced by the maximum powers of $M_H/M_W$---at
each order. By neglecting subleading corrections that involve $g$ or
the Yukawa couplings $g_f=\sqrt{2}m_f/v$, we expect to obtain
a good approximation to the full
result provided that $M_W/M_H\propto gv/M_H\ll 1$ and $m_f/M_H\ll 1$.
Because of the
high mass of the $t$ quark \cite{cdf},
it is interesting
to test the accuracy of the approximation.
In Fig.\ \ref{equiv_test}, we compare the
${\rm O}\left(G_FM_H^2\right)$ correction
to $\Gamma\left(H\rightarrow t\bar t\,\right)$,
already shown in Fig.\ \ref{hff_ratio}, to the full
one-loop electroweak correction including the effects of fermions
\cite{hff}. The full correction was evaluated in the on-shell renormalization
scheme using $m_t=174\ {\rm GeV}$ \cite{cdf}.
We see that the ${\rm O}\left(G_FM_H^2\right)$ term underestimates the full
one-loop electroweak correction term by 32\% (24\%) at
$M_H=500 \ {\rm GeV}$ (1~TeV). To check that the difference arises primarily
from the inclusion of the top quark in the full calculation,
and not from the supposedly small
contributions from the gauge sector of the SM, that is, from a failure of the
equivalence theorem, we have carried out a complete one-loop calculation
using the equivalence theorem with the gauge couplings set to zero \cite{ldkr},
but the top-quark Yukawa coupling retained.
The extra contributions are ${\rm O}\left(G_Fm_t^2\right)$, and are
independent of those considered above. As shown in Fig.\ \ref{equiv_test},
the result of the calculation reproduces the full one-loop electroweak result
very well. The result obtained using the equivalence theorem with $g_t\not=0$
is only 3.9\% (1.8\%) larger than the full
electroweak one-loop term at $M_H=500\ {\rm GeV}$ (1 TeV) for $m_t=174\
{\rm GeV}$. The use of the equivalence theorem therefore gives a quite
accurate approximation to the full theory, even for the rather low
values of $M_H$ with which we are concerned. The small residual differences
away from the decay threshold at $M_H=2m_t$
can be accounted for by the transverse gauge couplings,
the nonzero masses of the
$W$ and $Z$ bosons, and the finite masses and Yukawa couplings for the
remaining
fermions. The extra structure close to the
threshold is the result of
virtual-photon exchange in QED.
This generates a Coulomb singularity
and a correction that behaves near threshold as
$1+\alpha_{\rm em}Q_t^2[(\pi/2\beta)+{\rm O}(1)]$,
where $Q_t$ and $\beta$ are the  top-quark electric charge and velocity;
see left end of the dashed line in Fig.\ \ref{equiv_test}.

In Fig.\ \ref{QCD} we compare the electroweak and QCD corrections.
The latter were calculated with the asymptotic scale parameter of QCD
adjusted to give $\alpha_{\rm s}(M_Z)=0.118$ \cite{blo}. The
one-loop QCD correction to  $\Gamma\left(H\rightarrow t\bar{t}\right)$
in Fig.\ \ref{QCD}(a)
shows the expected color-Coulomb threshold singularity, with
$\alpha_{\rm em}Q_t^2$
replaced in the expression above by $(4/3)\alpha_{\rm s}(M_H)$.
This singularity is associated with the nonrelativistic motion of the
quarks. The set of correction terms in powers of $(2\pi\alpha_{\rm s}/3\beta)$
corresponds to the expansion of a Coulomb wave function at zero quark
separation. For $M_H\gg2m_t$, the one-loop QCD correction is negative, with
a magnitude which increases logarithmically. The two-loop QCD corrections to
$\Gamma\left(H\rightarrow t\bar t\,\right)$ are unknown.
They are expected to be large close to the $t\bar t$ production threshold,
at $M_H\gtrsim2m_t$.
At $M_H\gg2m_t$,
the potentially large logarithmic contributions in all higher orders
can be resummed by using the top-quark $\overline{\rm MS}$ mass
evaluated at the scale $M_H$,
and the residual corrections should be small (see the discussion given
below for the $b\bar{b}$ decay).

In Fig.\ \ref{QCD}(b), we repeat the comparison of
Fig.\ \ref{QCD}(a) for the case $H\rightarrow b\bar{b}$ assuming $m_b=4.72\
{\rm GeV}$ \cite{dom}.
The difference between the full one-loop electroweak correction and
the ${\rm O}\left(G_FM_H^2\right)$ result is again accounted for at
large values of $M_H$ by the omission of top-quark effects in the latter.
The spikes in the full correction at $M_H=2M_W$ and $2M_Z$ originate in
threshold singularities of the Higgs-boson wave-function renormalization.
The dent at $M_H=2m_t$ is not accompanied by such a divergence.
These features may be understood as artifacts of the underlying approximation
of treating the unstable Higgs boson as an asymptotic state.
The QCD corrections  are calculated to
${\rm O}\left(\alpha_{\rm s}^2\right)$
in the $\overline{\rm MS}$ scheme \cite{bra,gor}; see Eq.~(37) of \cite{mod}.
When the quark pole mass is used as a basic parameter,
the largest part of the one-loop QCD correction comes from a large
logarithmic term $-(4\alpha_{\rm s}/\pi)\ln\left(M_H/m_b\right)$ \cite{bra}.
In general, the large logarithms are of the form
$(\alpha_s/\pi)^n\ln^m(M_H/m_b)$,
with $n\ge m$. By
exploiting renormalization-group techniques, these logarithms may be
absorbed completely into the $\overline{\rm MS}$ quark mass, $m_b(\mu)$,
evaluated at the scale $\mu=M_H$ \cite{kni}.
The logarithms are resummed to all orders, and the remaining perturbative
expansion converges more rapidly.
The {\it offset} seen in the Fig.\ \ref{QCD}(b), with the QCD-corrected
decay considerably below the Born decay rate, results mainly from the use
of $m_b(M_H)$ instead of $m_b$ in the prefactor in Eq.\ (\ref{eqborn}).
While the effect is large, it is controlled by the resummation.
The remaining part of the QCD correction at two loops is rather small.

\subsection{Handling a nonperturbative Higgs}

One would like to be able to describe the Higgs-boson decay to fermions
phenomenologically even in the case of a strongly interacting Higgs sector.
In the case of the QCD corrections discussed above, the renormalization
group provided a physical principle that could be used to motivate
and organize a resummation of higher-order effects to obtain a
controlled final expression, even though the corrections could be
large when viewed order-by-order in $\alpha_{\rm s}$.
We do not know of a similar physical organizing principle
to use in the resummation of higher-order corrections in $\lambda$,
especially as the leading corrections only depend on one energy scale, namely
the mass of the Higgs boson.  Any
summation of the perturbation series will therefore be speculative, and
will necessarily be based
on the mathematical structure of the series rather a physical argument.
We note in this connection that the correction terms in Eq.\ (\ref{eqexp})
alternate in sign.
This suggests that Pad\'e summation of the series might be reasonable.
In particular, we can use the two-loop information given in
Eq.\ (\ref{eqexp}) to rewrite the perturbative series
using a [1,1] Pad\'e approximant
\cite{bender}, that is, as a ratio of two first-degree polynomials
in $\lambda$, with the coefficients adjusted to fit the expansion
in Eq.\ (\ref{eqexp}) through order $\lambda^2$. We obtain

\begin{equation}
{{Z_H}\over {Z_w}} = 1 + {{(a_w-a_H)\hat{\lambda} }\over
        {1 - [{{(b_w-b_H-a_wa_h+a_H^2)}/ {(a_w-a_H)}}]\hat{\lambda} }}\,.
\label{eqpade}
\end{equation}

In Fig.\ \ref{Pade}, we compare the Pad\'e-summed correction factor,
Eq.\ (\ref{eqpade}), with the earlier results from Fig.\ \ref{hff_ratio}
or Eqs.\ (\ref{eqres}) and (\ref{eqexp}). The result suggests
that the leading electroweak corrections to $\Gamma\left(H\rightarrow f\bar{f}
\right)$ will be quite small
even for values of $M_H\gtrsim 1.5$ TeV.
How far the result can be trusted is a matter of speculation.
A similar Pad\'e summation of the partial-wave scattering amplitudes for
$W_L^\pm,\ Z_L,\ H$ scattering turns out to give a fairly good prediction
for the two-loop contribution in terms of the zero- and one-loop terms
\cite{thesis},
so the method may be more reliable in this rather similar case than our
limited input information would suggest. If so, the leading electroweak
corrections to $H\rightarrow f\bar{f}$
in powers of $\lambda$ or $G_FM_H^2$ will be negligible, when resummed,
relative to the corrections introduced by the Yukawa
couplings and QCD. Only experimental results or reliable nonperturbative
calculations can resolve this speculation.

We note in this connection that recent lattice simulations of certain
Yukawa models for the interaction of the Higgs boson
with mirror or reduced staggered fermions suggest that
the Yukawa couplings cannot be strong, unless the regularization scale is
unacceptably low \cite{fri}.

\section{CONCLUSIONS}

In summary, we have calculated the leading two-loop electroweak
corrections to the fermionic decay rates of a high-mass Higgs boson
in the SM, which are of ${\rm O}\left(G_F^2M_H^4\right)$. The corrections
are negative and exceed the positive ${\rm O}\left(G_FM_H^2\right)$ one-loop
corrections in magnitude for $M_H>1114\ {\rm GeV}$.
For larger values of $M_H$, the perturbation series is clearly unreliable,
and the theory becomes effectively strongly interacting.
We conclude, given the
lack of a physical principle that would allow a convincing resummation
of the perturbation expansion, that a value $M_H\sim 1100\ {\rm GeV}$
has to be considered as a theoretical
upper bound on $M_H$ beyond which the fermionic decay width of the Higgs cannot
be calculated perturbatively.
This result  is independent of speculations regarding
the energy scale up to which the SM is valid, as the center-of-mass energy in
the Higgs decay is fixed,
$\sqrt{s}=M_H$.
However, there is indication that high-energy interactions in the
Higgs sector of the SM can become effectively strong, and are not usefully
calculable in perturbation theory, for even smaller Higgs mass.
For
scattering processes with $\sqrt{s}\sim 5$~TeV, the critical value for $M_H$
is about
$380$ GeV, and for scattering at GUT energies the critical value is less than
$160$ GeV \cite{unit}.
Clearly, the present-day precision
tests of the gauge sector of the SM are not affected by such nonperturbative
effects.

\acknowledgments
One of us (BAK) would like to express his gratitude to the Physics
Department of UW-Madison for supporting his visit, during which part of
this work was carried out, and for the great hospitality extended to him.
This work was supported in part by the U.S. Department of Energy under
Contract No.\ AC02--76ER00881.

\begin{figure}
\caption{
The two-loop diagrams that contribute to the wave-function
renormalization constants $Z_H$ and $Z_w$ at ${\rm O}\left(G_F^2M_H^4\right)$
through derivatives of the
self-energy functions $\Pi^0_H$ and $\Pi^0_w$, Eq.\ (\protect\ref{eqrenorm}).
Heavy (light) lines represent Higgs ($w^\pm$ and $z$) bosons.
The statistical weights of the
diagrams are not shown, but may be read off from Eqs.\ (11) and (12) in
\protect\cite{mah}. See Appendix A in \protect\cite{mah} for the results.}
\label{diagrams}
\end{figure}

\begin{figure}
\caption{
Complete ${\rm O}\left(G_FM_H^2\right)$ and
${\rm O}\left(G_F^2M_H^4\right)$ correction factors
for $\Gamma\left(H\rightarrow f\bar f\,\right)$
for
$100 \ {\rm GeV}$ $\leq$ $ M_H$ $\leq 1700 \ {\rm GeV}$.
These corrections are universal, i.e., they are independent of the flavor of
the final-state fermions.
In each order, the expanded result given in Eq.\ (\protect\ref{eqexp})
is compared
to the calculation where the one-particle-reducible Higgs-boson
self-energy diagrams are resummed as shown in Eq.\ (\protect\ref{eqres}).
The two-loop correction cancels the one-loop correction
at $M_H=1114 \ {\rm GeV}$ and is twice as large as the latter,
with an opposite sign, at $M_H=1575 \ {\rm GeV}$.
}
\label{hff_ratio}
\end{figure}

\begin{figure}
\caption{
Comparison of the one-loop
results for the ratio $\Gamma\left(H\rightarrow t\bar{t}\right)/
\Gamma_B\left(H\rightarrow t\bar{t}\right)$ obtained
in various approximations with the full one-loop electroweak result
$(g_1,\ g_2,\ g_t,\ g_b \not = 0)$.
The solid curve (EQT) gives the result obtained
using the equivalence
theorem with vanishing gauge couplings $(g_1,\ g_2 \not = 0)$  and a nonzero
top-quark Yukawa  coupling corresponding $g_t$ to
$m_t=174\ {\rm GeV}$. The dot-dashed curve shows the ${\rm O}(\lambda)\hat{=}
{\rm O}\left(G_FM_H^2
\right)$ correction from Fig.\ \protect\ref{hff_ratio},
and it is equivalent to an EQT curve with $m_t$=0.
}
\label{equiv_test}
\end{figure}

\begin{figure}
\caption{
Electroweak and QCD correction factors for
(a) $\Gamma\left(H\rightarrow t\bar t\,\right)$,
and (b) $\Gamma\left(H\rightarrow b\bar b\,\right)$ as a function of $M_H$:
universal ${\rm O}\left(G_FM_H^2\right)$ term without resummation (solid line);
full one-loop electroweak corrections (dashed line);
QCD corrections (dot-dashed line);
and universal ${\rm O}\left(G_FM_H^2\right)$ plus
${\rm O}\left(G_F^2M_H^4\right)$ terms
without resummation (solid line).
The QCD corrections are evaluated in
(a) to ${\rm O}(\alpha_{\rm s})$ in the on-mass-shell scheme,
and in (b) to ${\rm O}\left(\alpha_{\rm s}^2\right)$
in the $\overline{\rm MS}$ scheme.
The pole-mass values $m_t=174 \ {\rm GeV}$ and $m_b=4.72 \ {\rm GeV}$
\protect\cite{dom} are used,
and the asymptotic scale parameter of QCD is
adjusted so that $\alpha_{\rm s}(M_Z)=0.118$
\protect\cite{blo}.
Note the different scales used in the two plots.
}
\label{QCD}
\end{figure}

\begin{figure}
\caption{
Comparison of the Pad\'e-resummed correction for
$\Gamma\left(H\rightarrow f\bar f\,\right)$
with the universal electroweak correction factors calculated
to ${\rm O}\left(G_FM_H^2\right)$ and to
${\rm O}\left(G_F^2M_H^4\right)$.
}
\label{Pade}
\end{figure}


\begin{references}
\bibitem{adrian}
The results in \cite{prl} and \cite{ghi} have both been revised, and now
agree completely. We would like to thank Dr.\ A.\ Ghinculov for his
cooperation comparing the two independent calculations.
\bibitem{breaking}The SO(3) symmetry of the $(w^\pm,z)$ Goldstone bosons
is broken by the
interactions of the $w^\pm$ and $z$ with fermions, but this does not affect
the dominant two-loop corrections. The complete renormalization scheme
for use with the Goldstone-boson equivalence theorem is discussed in
\cite{ldkr}.
\end{references}
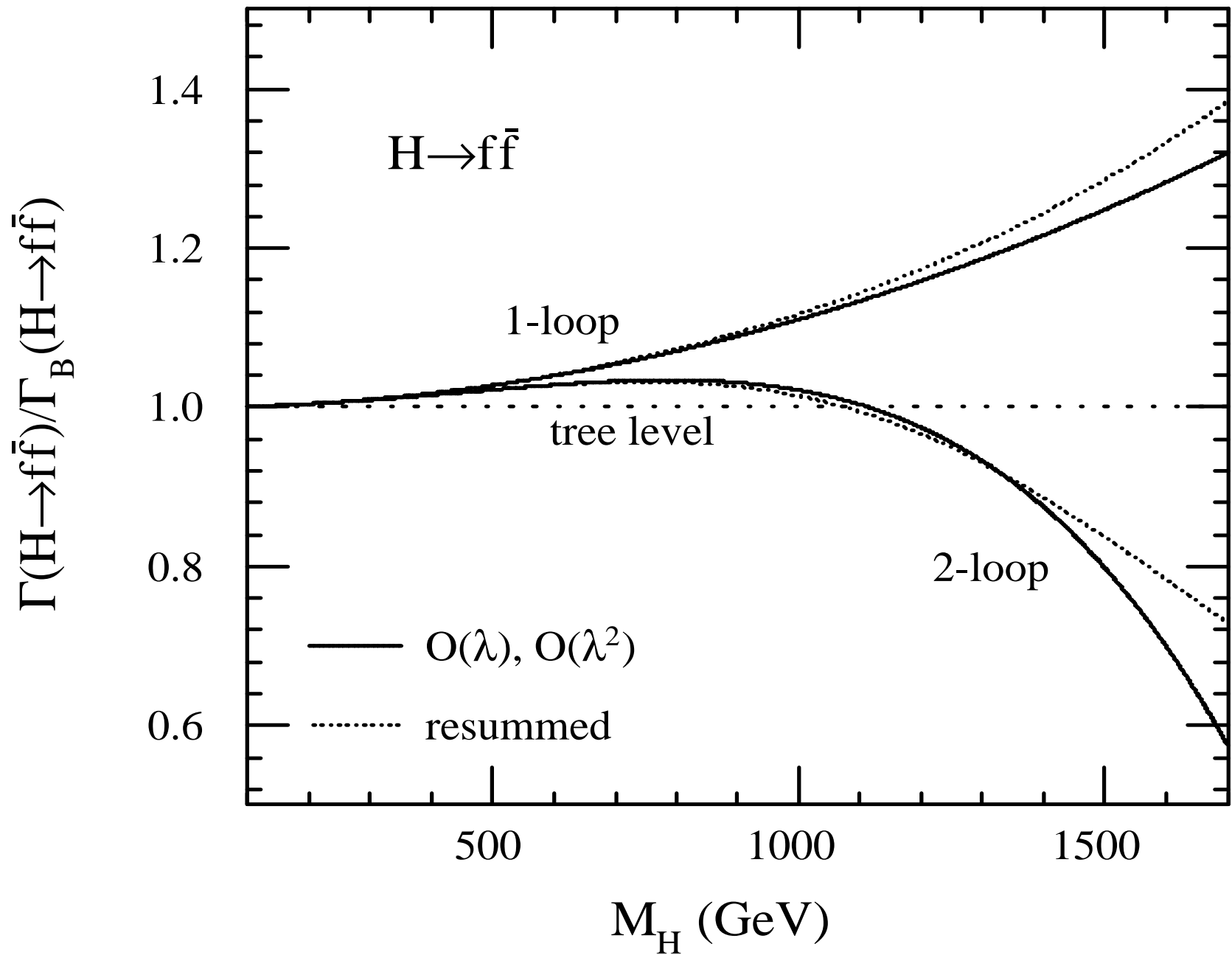
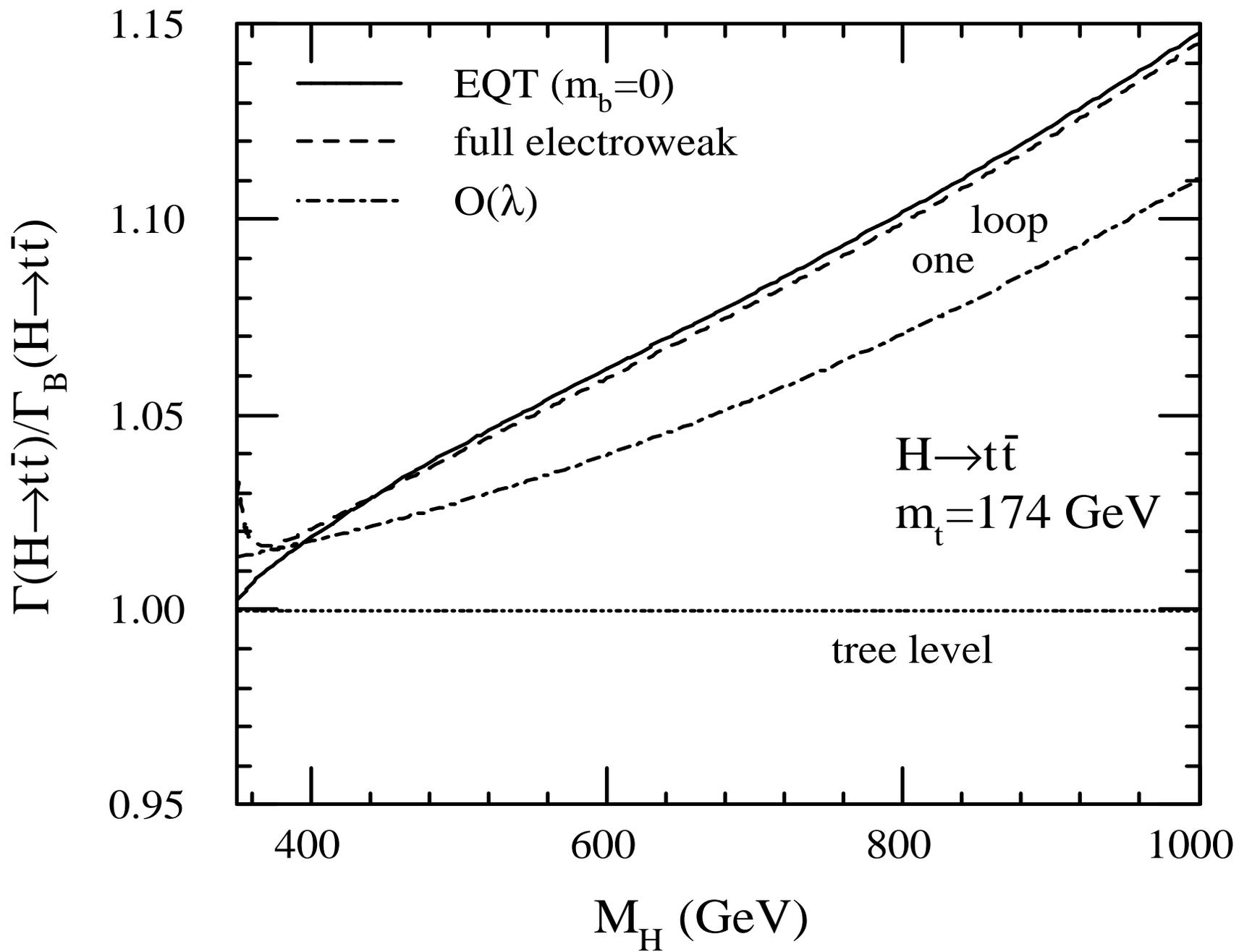
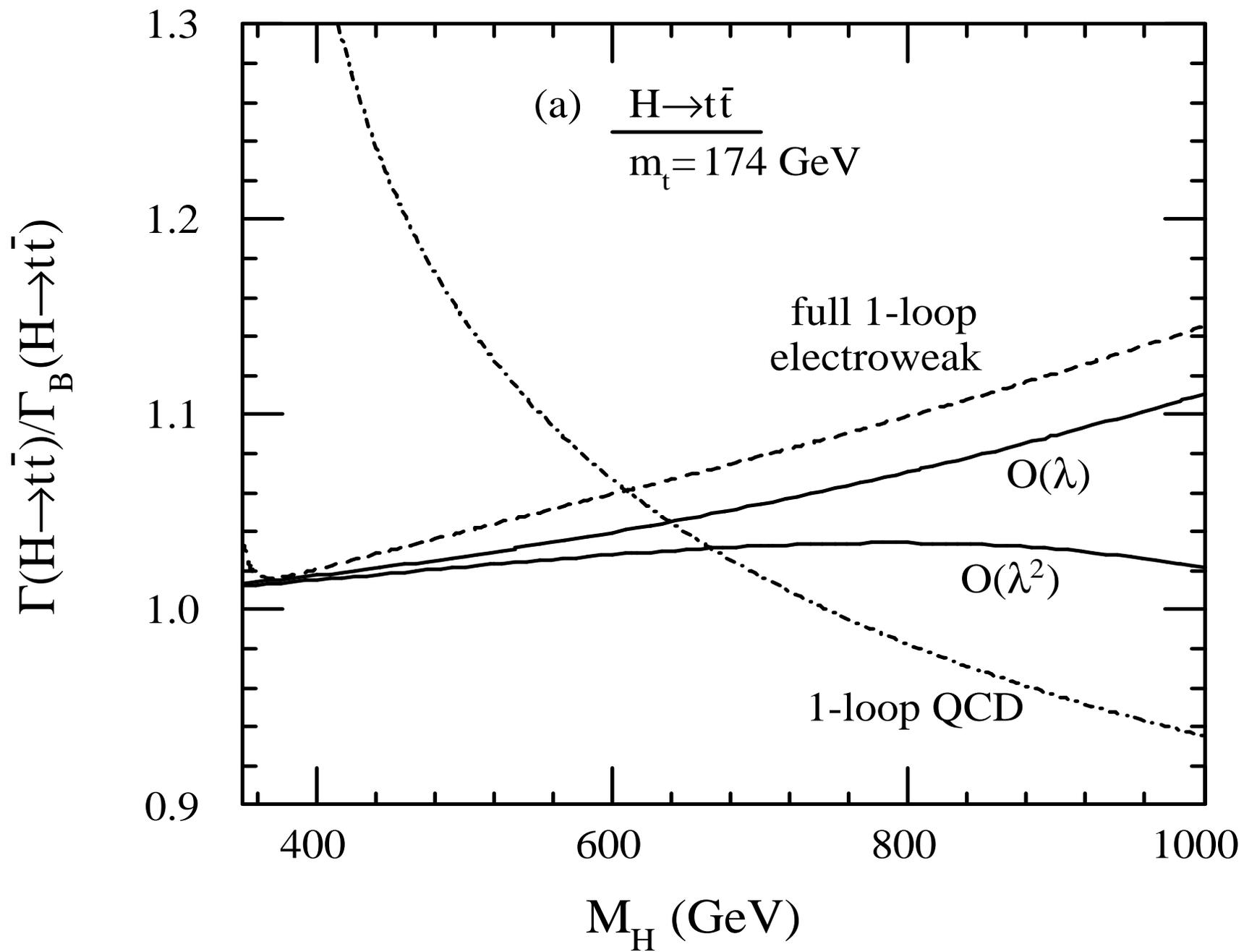
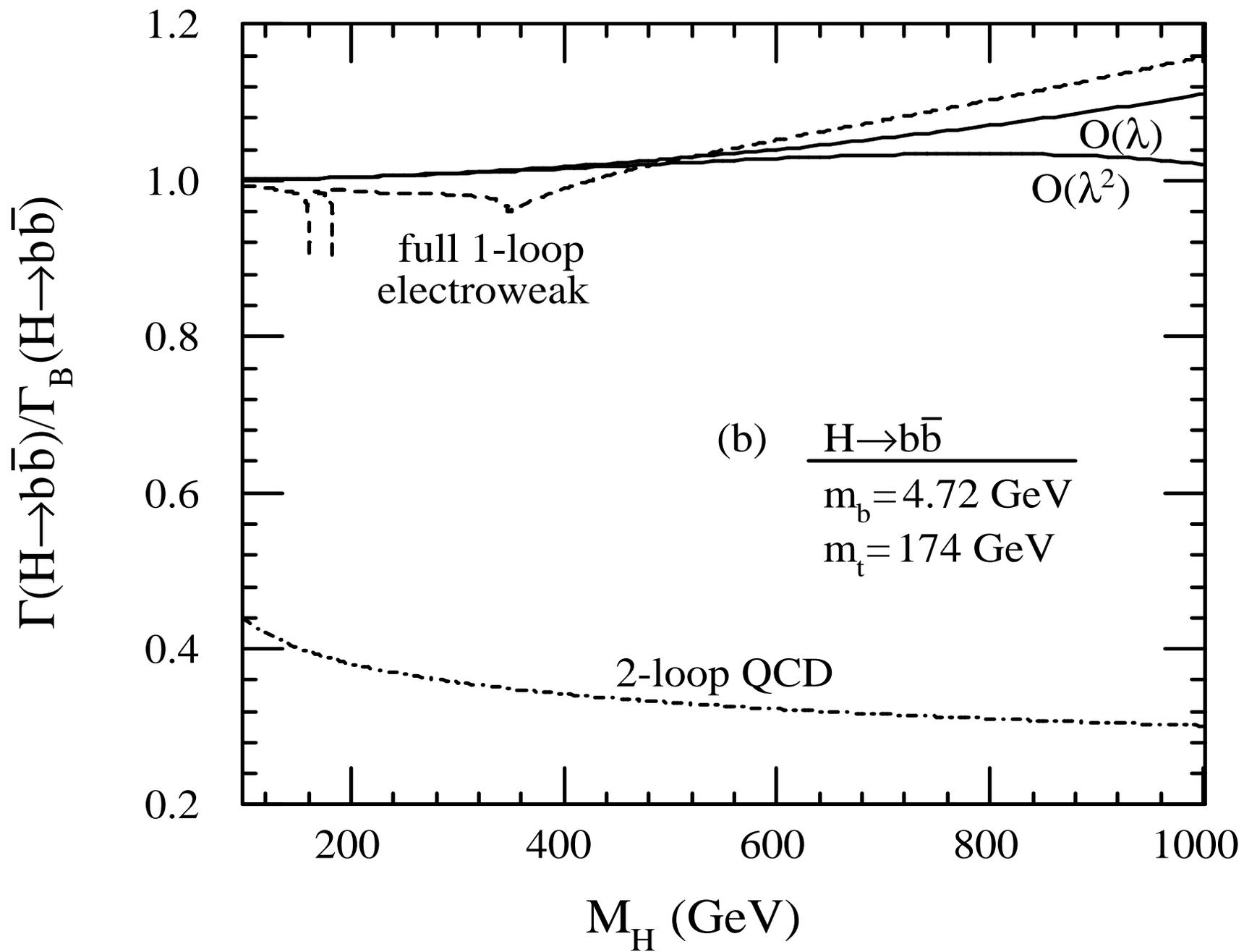
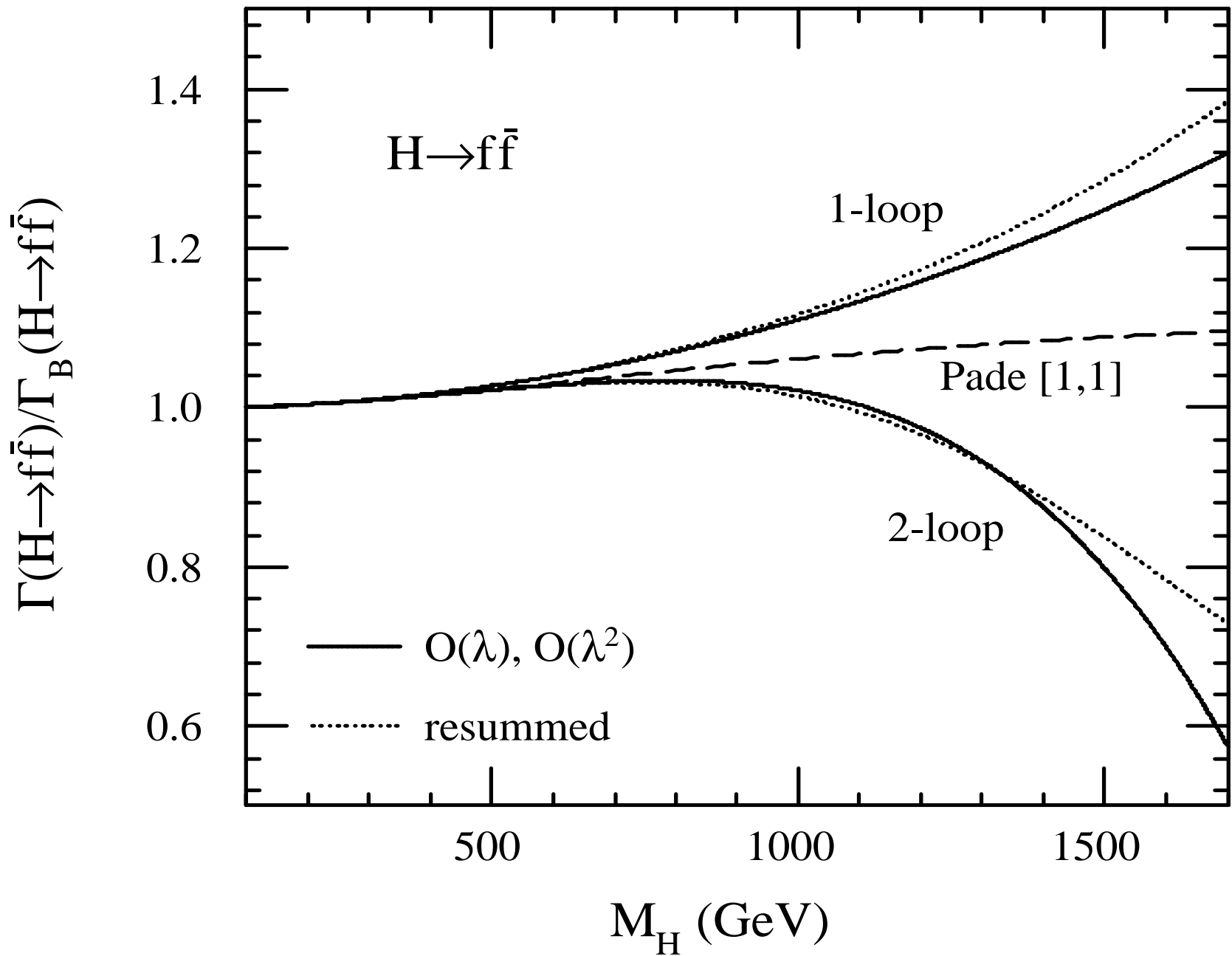
\end{document}